# Soft x-rays with Orbital Angular Momentum for resonant scattering experiments at the SOLEIL synchrotron

Pietro Carrara,[a,*] Franck Fortuna,[b] Renaud Delaunay,[c] Joan Vila-Comamala,[d] Benedikt Rösner,[d] Christian David,[d] Stefania Pizzini,[e] Clément Fourniols,[f] Laurent Vila,[f] Matteo Pancaldi,[g] Carlo Spezzani,[g] Flavio Capotondi,[g] Pierre Nonnon,[h,i] Mauro Fanciulli,[h,i] Thierry Ruchon,[i] Nicolas Jaouen,[j] Horia Popescu,[j,*] Maurizio Sacchi,[a,j,*]

[a] Sorbonne Université, CNRS, Institut des NanoSciences de Paris, INSP, 75005 Paris, France
[b] Université Paris-Saclay, CNRS, Institut des Sciences Moléculaires d'Orsay, 91405, Orsay, France
[c] Sorbonne Université, CNRS, Laboratoire de Chimie Physique—Matière et Rayonnement, LCPMR, 75005 Paris, France
[d] Center for Photon Science, Paul Scherrer Institute, 5232 Villigen PSI, Switzerland
[e] Université Grenoble Alpes, CNRS, Institut Néel, 38042 Grenoble, France
[f] Université Grenoble Alpes, CNRS, CEA, Grenoble INP, IRIG-SPINTEC, 38000 Grenoble, France
[g] Elettra-Sincrotrone Trieste, Strada Statale 14 km 163,5 in Area Science Park, 34012 Basovizza, Trieste, Italy
[h] CY Cergy Paris Université, CEA, LIDYL, 91191 Gif-sur-Yvette, France
[i] Université Paris-Saclay, CEA, LIDYL, 91191 Gif-sur-Yvette, France
[j] Synchrotron SOLEIL, Saint-Aubin, Boite Postale 48, 91192 Gif-sur-Yvette, France

* Correspondence: pietro.carrara@insp.jussieu.fr, horia.popescu@synchrotron-soleil.fr, maurizio.sacchi@synchrotron-soleil.fr

**Abstract** The paper presents a comprehensive description of a new setup implemented and commissioned at the SEXTANTS beamline of the SOLEIL synchrotron for absorption and scattering experiments with x-ray beams carrying an orbital angular momentum, also known as *twisted* x-ray beams. Two alternative methods have been implemented, based on the use of either spiral zone plates or fork gratings devices, and we show how they can be used for both defining and assessing the orbital angular momentum of an x-ray beam. We show also how multiple devices can be used in sequence to define an *integer arithmetic* of the orbital angular momentum of the final x-ray beam. Finally, we report the results of the first resonant scattering pilot experiments in transmission and reflection mode, intended to assess the feasibility of future users' measurements. The availability of twisted soft x-rays complements the range of experimental techniques in elastic, resonant and coherent scattering available at the SEXTANTS beamline of the SOLEIL synchrotron.

## 1. Introduction

In addition to the spin angular momentum (SAM) that originates from the light polarization, photon beams can carry also an orbital angular momentum (OAM) of $L\hbar$ per photon, with $L \in \mathbb{Z}$, associated to an azimuthally varying phase term $\exp(iL\phi)$ in the Laguerre-Gaussian representation of the electromagnetic radiation (Allen *et al.*, 1992). In the visible and near-infrared (vis-IR) range of wavelengths, OAM laser beams, also called *twisted optical beams*, find applications in fields as varied as biology, telecommunication, imaging and quantum technologies (Shen *et al.*, 2019). Their capability to exert a mechanical torque has been exploited to create *optical spanners* for manipulating small particles (Simpson *et al.*, 1997; Ladavac & Grier 2004). The azimuthal phase dependence introduces a singularity on the propagation axis and a radial modulation of the intensity, which results in a ring-shaped beam. Such properties have been used to modify magnetic ordering (Fujita & Sato 2017), to improve the spatial resolution in microscopy (Tamburini *et al.*, 2006), and to enhance the edge sharpness in phase contrast imaging (Fürhapter *et al.*, 2005). More recently (Géneaux *et al.*, 2016; Gauthier *et al.*, 2017; Rebernik Ribič *et al.*, 2017), the generation of OAM beams at shorter wavelengths, from XUV to hard x-rays, is also finding an increasing number of applications, often based on extrapolations of previous work carried out in the vis-IR range. For instance, a recent study by Pancaldi *et al.* (2024) compared XUV ptychographic images of a reference sample obtained using beams with different OAM values, confirming that the attainable spatial resolution improves with $L$. Extending the use of OAM beams from the vis-IR (Tamburini *et al.*, 2006) to the XUV (Wang *et al.*, 2023; Pancaldi *et al.*, 2024) and x-rays opens new perspectives for high-resolution element-selective x-ray imaging.

X-ray radiation carrying orbital angular momentum (X-OAM) has received increasing interest over the last few years, both in terms of theoretical studies (Nazirkar *et al.*, 2024; Yan, & Geloni, 2023; Moghaddasi Fereidani *et al.*, 2025) and of practical applications in different fields of research (Fujita & Sato 2017; Loetgering *et al.*, 2020; Woods *et al.*, 2021; McCarter *et al.*, 2023). As with SAM, the handedness imposed by the OAM has been exploited to perform spectroscopic studies of chiral molecules with hard x-rays (Rouxel *et al.*, 2022) and of magnetic materials in the XUV range (Fanciulli *et al.*, 2022; Fanciulli *et al.*, 2025). In particular, the concept of magnetic helicoidal dichroism (MHD) has been introduced to describe the dependence of the resonant absorption and scattering from a magnetic material on the OAM value of the incident photon beam (Fanciulli *et al.*, 2021; Ruchon *et al.*, 2022; Luttmann *et al.*, 2025).

Here we present a new setup dedicated to absorption and scattering experiments with soft x-rays carrying OAM, which has been commissioned at the SEXTANTS beamline (Sacchi *et al.,* 2013a) of the SOLEIL synchrotron source. We implemented two ways of producing OAM beams, using either spiral zone plates (SZPs, Vila-Comamala *et al.,* 2014) or fork gratings (FGs, Lee *et al.,* 2019), and explored the possibility of generating OAM radiation directly from helical undulator sources (Sasaki et al., 2007; Sasaki & McNulty, 2008). Finally, we present the results of test measurements performed using these OAM beams.

## 2. Description of the experimental setup

The IRMA-2 endstation (Sacchi et al., 2013b) was installed at the end of the SEXTANTS elastic scattering branch (Sacchi et al., 2013a). Approximately 9 m upstream of the endstation, the beamline features an intermediate focal point where a pin-hole (10, 20 or 50 µm in diameter) can be inserted to create a secondary source, trimming the beam and ensuring a high degree of transverse coherence. Downstream of the secondary source, a Kirkpatrick-Baez (KB) refocusing optics with bendable mirrors, placed approximately 3 m before the IRMA-2 chamber, allows for either a large (≥ 2 mm) or focused (~40 µm × 30 µm, FWHM horizontal × vertical) beam to illuminate the SZP or the FG, respectively.

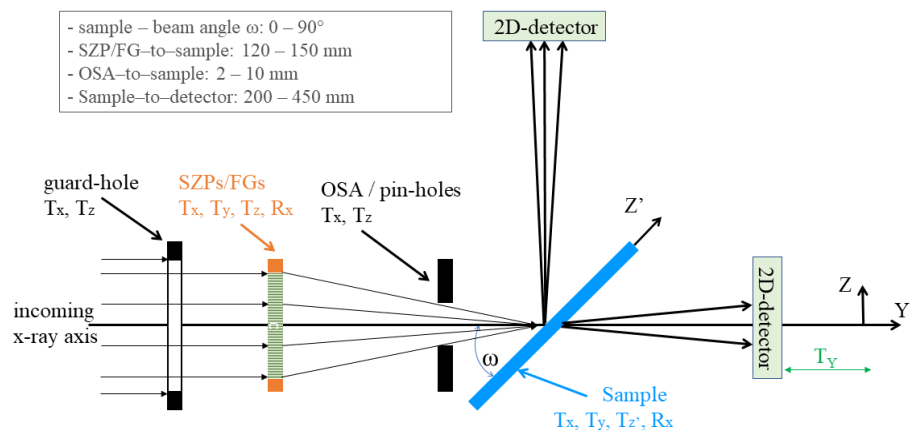

**Figure 1**
Side-view sketch (not to scale) of the experimental setup inside the IRMA-2 chamber.

The experimental setup, sketched in Fig.1, consists of the following elements:

- a fixed 2.5 mm circular aperture for skimming the beam before the experimental chamber; this is especially important under defocused KB conditions for SZP illumination.

- encoded XYZ-tables driven by piezoelectric motors, with additional $R_X$ manual alignment, hosting the SZPs and/or the FGs.

- piezo-driven XYZ-tables for aligning a set of order-sorting apertures (OSA) from 20 µm to 1 mm in size.

- high-precision $XYZR_X$ tables driven by stepper motors for aligning samples (or a second set of FGs), with translation ranges ≥ 25 mm.

- A long-travel translator fixed on a CF-150 flange carries the in-vacuum CCD camera (PI-MTE with 2048 × 2048 pixels, 13.5×13.5 µm² in size) equipped with a light-tight Zr filter. The sample to CCD distance can be varied between 200 and 450 mm, allowing for a trade-off between the angular range covered by the detector and the angular resolution of a single pixel, depending on the specific needs of each experiment. A beam-stop can be positioned over the detector active area using two piezo-driven translation tables attached to the CCD frame.

The detection assembly can be mounted on the IRMA-2 chamber either in horizontal or in vertical position, to measure in transmission or reflection geometry, respectively. Positioning the CCD at intermediate angles (ranges 30±10° and 60±10°) is also possible using CF-100 mounting flanges (Sacchi *et al.,* 2013b), but keeping a fixed distance of 450 mm from the sample. For alignment purposes, the radiation intensity, either transmitted or reflected, can also be measured by retractable photodiodes placed in between the sample and the CCD.

With the exception of results shown in **Appendix A**, the test measurements reported here were performed in transmission mode.

## 3. Spiral zone plates

SZPs are diffractive focusing elements that in addition to producing a highly demagnified and convergent beam also impart OAM to the x-rays. The **L** value of the outgoing beam is determined by the number *s* of spiraling branches (Vila-Comamala *et al.,* 2014), with the sign of *s* defining whether the spirals turn clockwise (CW) or counter-clockwise (CCW) with respect to the wavevector of the incoming photons. SZP devices have been tested with x-rays for a few years already, in experiments performed at synchrotrons (Cojoc *et al.*, 2006; Kohmura *et al.*, 2020; Rouxel *et al.*, 2022) and free-electron lasers (Fanciulli *et al.*, 2022).

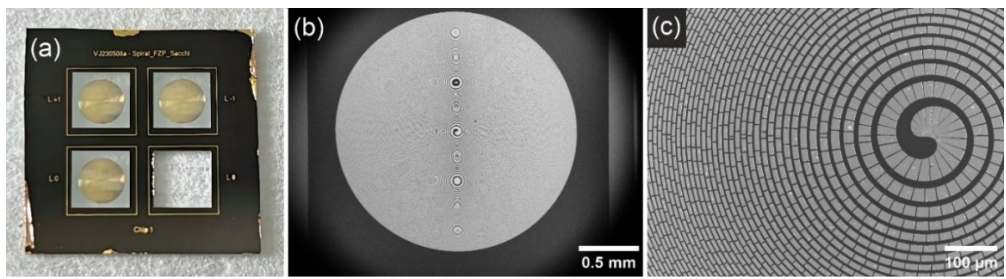

**Figure 2**

(a) Picture of a 10 mm Si frame hosting three SZPs with *s* = -1, 0 and +1. (b,c) Scanning electron microscopy (SEM) images of the *s* = +1 SZP with different magnifications [scalebars are 0.5 mm in (b) and 0.1 mm in (c)].

SZPs with *s* = 0, ±1, ±2, ±3 were manufactured at the Laboratory for X-ray Nanoscience and Technologies of the Paul Scherrer Institut (Villigen, Switzerland) by electron-beam lithography, Au thermal evaporation and lift-off process on silicon nitride ($Si_3N_4$) membranes supported by a Si frame (Fig. 2). The Au layer thickness of 120 nm was optimized for high efficiency at around 150 eV (Gd and Tb $N_{4,5}$ edges). Each SZP has a diameter of 2 mm, with an outermost zone width of ~600 nm, and a duty cycle of ~2/3 to improve the 1st order over 0th order efficiency. The expected minimum beam waist at 150 eV is of the order of 1 µm.

The parameters common to all SZPs (i.e., apart from their *s* value) are summarized in Table I.

| | |
|---|---|
| Si frame size | 10×10 mm$^2$ |
| $Si_3N_4$ membrane size | 2.5×2.5 mm$^2$ |
| $Si_3N_4$ membrane thickness | 50 nm |
| Cr coating | 5 nm |
| Au film thickness | 120 nm |
| SZP diameter | 2 mm |
| SZP smaller outer zone | ~600 nm |
| SZP duty cycle | ~2/3 |
| Focal distance @150eV | ~145 mm |
| Spot size FWHM | ~1 µm |

**Table I.** Design parameters common to all SZPs.

One of a set of circular apertures (diameter 10, 20 or 50 µm) was inserted at the intermediate focal point of the beamline to generate a secondary source with large transverse coherence, and the bendable KB mirrors are adjusted to have a collimated beam configuration. A fixed 2.5 mm aperture placed between the KB and the SZPs ensures a good match of the beam and SZP sizes. Higher diffraction orders are spatially filtered by an order sorting aperture (OSA) placed before the sample at an adjustable distance (1-10 mm). OSAs of different sizes, from 20 µm to 1 mm, are available on the same support and can be interchanged in situ. Test samples are mounted on the $XYZR_X$ stage (Fig. 1), and measurements are performed using the in vacuum CCD. The spot size produced by the SZP with *s* = +1 at the sample position was estimated by knife-edge XZ scans as a function of the SZP coordinate along Y. After optimizing the KB mirror parameters to minimize astigmatism, the smallest spot at 150 eV photon energy was always approximately 3 µm FWHM, a value larger than expected, probably due to residual imperfections in the beam wavefront.

To test the azimuthal dependence of the phase generated by the SZP we made use of the interference between the diffracted beam, carrying the desired **L** value, and the beam transmitted through the SZP and the OSA, which retains the initial **L** = 0 value. Fig. 3 shows the calculated (bottom) and measured

(top) interference patterns for different SZPs ($s = 0, +1, +2$, and $-3$). The $s = 0$ zone-plate produces a ring-shaped intensity pattern, due to the variation of optical path between transmitted and diffracted waves. For $s = +1$, the additional azimuthal phase modulation leads to a CW spiraling intensity, which becomes a CW two-spirals for $s = +2$ and a CCW three-spirals for $s = -3$.

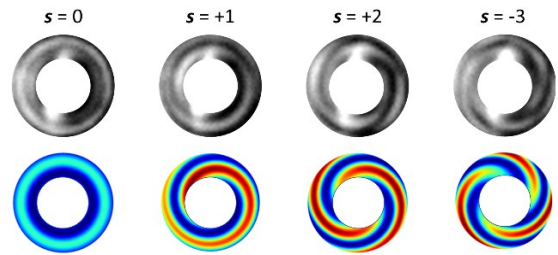

**Figure 3**

Comparison of experimental (top) and calculated (bottom) intensity modulations arising from the interference between the wave diffracted by a SZP with $s$ branches ($s = 0, 1, 2, -3$), having $\mathbf{L} = s$, and the wave transmitted through the same SZP, having always $\mathbf{L} = 0$.

Fig. 4 shows the intensity of the 155 eV radiation with $\mathbf{L} = +1$ transmitted through a 50 nm thick $Co_{91}Tb_9$ alloy sample and collected on the CCD at its closest position, i.e. at a sample–CCD distance of ~200 mm. The central high intensity disk (saturated scale) corresponds to the diverging beam from the SZP. Its diameter (~2.8 mm) is consistent with the 2 mm size of the SZP positioned 145 mm upstream of the sample. The weaker intensity modulations (speckle) observed at larger angles are due to resonant scattering at the Tb-$N_{4,5}$ edge from the up/down magnetization in the meander domains of the perpendicularly magnetized $Co_{91}Tb_9$ sample.

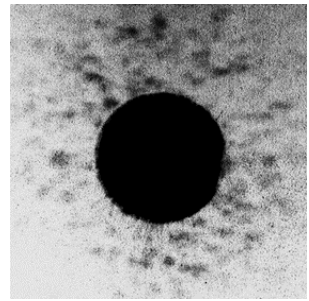

**Figure 4**

Intensity transmitted through a 50 nm thick $Co_{91}Tb_9$ alloy sample using a SZP with $s = +1$ at 155 eV (Tb-$N_{4,5}$ edge). The image (512 pixels, or ~6.9 mm, wide) was collected with the CCD detector ~200 mm behind the sample. The intense central spot corresponds to the diverging beam diffracted by the SZP and transmitted through the sample. Speckles observed at larger angles are due to the resonant coherent scattering from the meandric magnetic domain structure.

The SZPs were used also for test experiments in reflectivity mode at the Gd-N4,5 edge. An example is reported in **Appendix A**.

## 4. Fork gratings

FGs (see Fig. 5) are diffractive elements frequently employed in the visible range for producing OAM beams (Janicijevic & Topuzoski, 2008; Hu *et al.*, 2022), and extensions to the x-ray range have already been proposed (Lee *et al.*, 2019). The OAM value **L** of the produced beam is given by the product of the number *f* of additional lines contained in the fork (also referred to as the topological charge of the FG) and the diffraction order *n*, i.e. **L** = *f* × *n*. Note that *f* can be positive or negative, corresponding to the fork pointing up of down, so that the sign of **L** is defined by the sign of both *f* and *n*. We will only consider positive *f* values in the following.

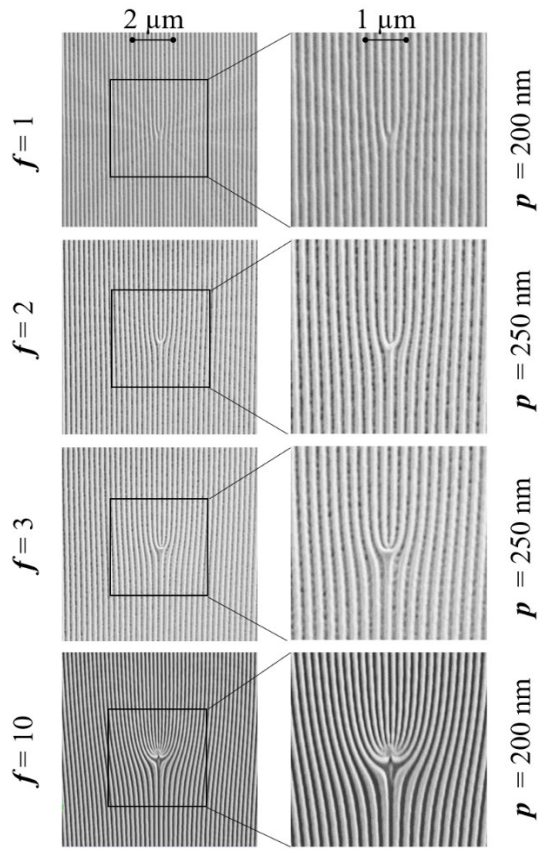

**Figure 5**

Left: SEM images of four FGs with different *f* and *p* values. Right: blowup of the central part of the FGs.

To test their potential as an alternative to SZPs in the soft x-ray range, we first used a set of small (10 × 10 µm²) FGs prepared by focused ion beam (FIB) etching of a continuous Au layer deposited on a 100 nm thick silicon nitride membrane. The grating period *p* is either 200, 250 or 600 nm, and the parameter *f* varies between 1 and 10, see Fig. 5. The thickness of the Au layer is 140 nm, optimized for efficient $n = \pm 1$ diffraction over the 140 – 180 eV energy range, with an angular separation of ~0.7° between 0th and 1st orders for $p$ = 600 nm.

By optimizing the parameters of the bendable KB mirrors, the beam size at the CCD position was set to 40 × 30 µm². As a consequence, the diffraction orders are affected by scattering from the edges of the FGs, which is smaller than the beam. Although this configuration is not ideal for experiments, it does not affect the conclusions of this test work. The possibility of fabricating larger FGs by lithography and ion-beam etching has already been addressed (see **Appendix B**).

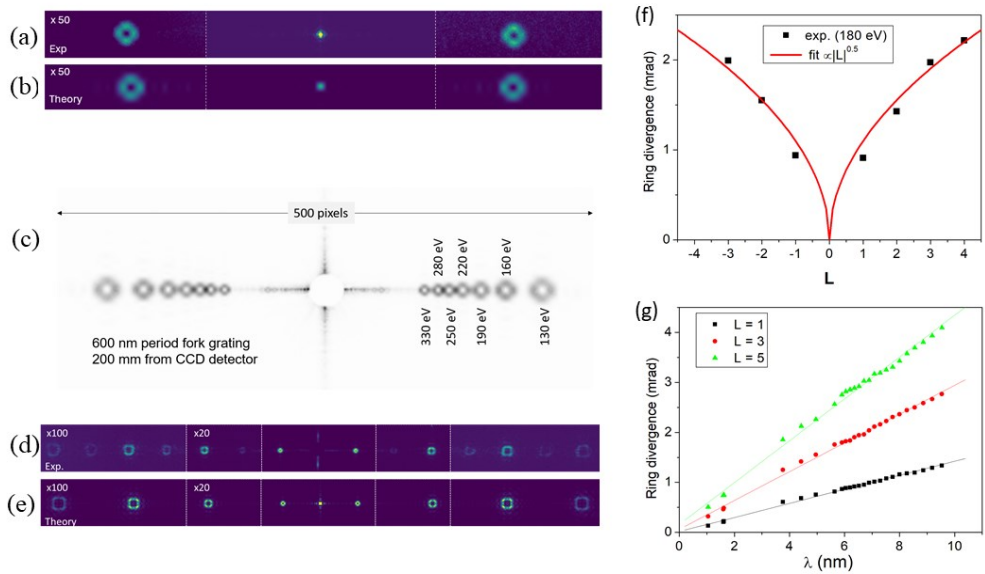

**Figure 6**

All reported results refer to a FG with $f$ = 1 and $p$ = 600 nm. (a) Intensity transmitted ($n$ = 0, central spot) and diffracted ($n$ = ±1, multiplied by 50) by the FG for a photon energy of 150 eV. (b) Numerical simulation of the results shown in (a), using the nominal FG parameters. (c) Accumulation of seven scattering patterns ($n$ = ±1) at varying photon energies over the 130-330 eV range, while keeping the positions of the FG and CCD detector fixed. (d) Diffraction orders up to $n$ = ±7 at a photon energy of 778 eV. The $n$ = 0 beam is blocked by a beam-stop. Relative to $n$ = 1, the intensity is multiplied by 20 for $n$ = 4, 5 and by 100 for $n$ = 6, 7. (e) Numerical simulation of the results shown in panel (d). (f) Points: experimental $\theta$ values for several diffraction orders at $\lambda \approx 6.9$ nm (180 eV). Line: fit with $|L|^{1/2}$ dependence. (g) Experimental $\theta$ values *vs.* $\lambda$ for three diffraction orders over the ~1–10 nm range, corresponding to the 130-1180 eV photon energy range.

Fig. 6(a) shows that the intensity of the $n = \pm 1$ diffraction peaks exhibits the ring shape expected for a beam with $\mathbf{L} \neq 0$. Numerical simulations of the beam propagation, performed using the open-source Python package *diffractio* (Sanchez Brea, 2019), show good agreement with the experimental results [Fig. 6(b)] in terms of angular position, shape and intensity of the diffracted orders. The intensity scale in Fig. 6(a) is multiplied by 50 for the diffracted orders $n = \pm 1$. The integrated intensity of the $n = \pm 1$ orders is about 9% of that measured for $n$ = 0, a ratio lower than the calculated 14.5% from data in Fig. 6(b). This is due to the size of the incoming beam being larger than the FG, the surrounding Au layer being partial transparent at 150 eV photon energy, but also to imperfections in the FG bar-shape and duty cycle, and to the possible presence of higher orders contamination from the x-ray beamline, all these sources contributing to the *n* = 0 intensity only.

Unlike SZPs, FGs do not act as focusing elements. Combined with the use of a 2D detector, this makes it easier to vary the photon energy without moving other mechanical parts (a lateral translation of the sample may still be required, depending on its size and homogeneity). Fig. 6(c) shows the *n* = ±1 diffracted intensity collected by the CCD detector kept at a fixed position for 7 different photon energies in the 130-330 eV range. By adjusting the CCD distance and accepting a lower efficiency of the FG, diffraction patterns were collected also at photon energies well above the optimized range, up to approximately 1200 eV. Fig. 6(d) shows an example measured at 778 eV, corresponding to the Co $L_3$ edge associated with a 2p-to-3d electron excitation (note that the intensity of the *n* ≠ 1 diffracted beams has been multiplied by either 20 or 100). Numerical simulations again reproduce the experimental results well [Fig. 6(e)]. In particular, the distortion of the ring shape due to scattering from the edges of the small grating area, especially pronounced at high diffraction orders, is correctly reproduced by considering in the simulations a square FG (10 µm) smaller than the incoming beam size (40 µm). The absence in the simulations of the faint even diffraction orders that can be observed in the measurements is due to the assumption of perfectly rectangular bars with exactly half-period width, which is not strictly fulfilled by the real grating. To each diffracted beam with OAM value **L** = *f* × *n* is associated a divergence *θ*, defined as the angle subtended by the ring of maximum intensity. This divergence is expected to be proportional to $\lambda(\mathbf{L})^{1/2}$, where λ is the x-ray wavelength (note that deviations from the 0.5 exponent have been reported in the optical range; Karimi *et al.*, 2007). Fig. 6(f) compares the measured divergence for several diffraction orders as a function of their expected **L** value with the square root dependence (red line). Fig. 6(g) shows the measured *θ*(λ) values for three diffraction orders, featuring a nearly linear dependence over the 130-1180 eV photon energy range.

In addition to confirming the expected radial distribution of the intensity [Fig. 6(a,d)], we verified also the azimuthal phase dependence for an **L** ≠ 0 beam diffracted by a FG by measuring the interference pattern generated by scattering through an opaque mask with five apertures (~1 µm in diameter) placed at the vertices of a pentagon [Fig. 7(a)], an approach similar to that reported by Hickmann *et al.* (2011). In Fig. 7(b), the interference patterns measured by illuminating the mask with diffraction orders *n* = -1, 0, +1 of the FG (*f* = 1) are compared with the far-field interference patterns calculated for incident beams with **L** = -1, 0, +1. The good agreement between experiments (*vs.* *n*) and calculations (*vs.* **L**) confirms that the beams generated by the FG possess the expected azimuthal dependence of the phase.

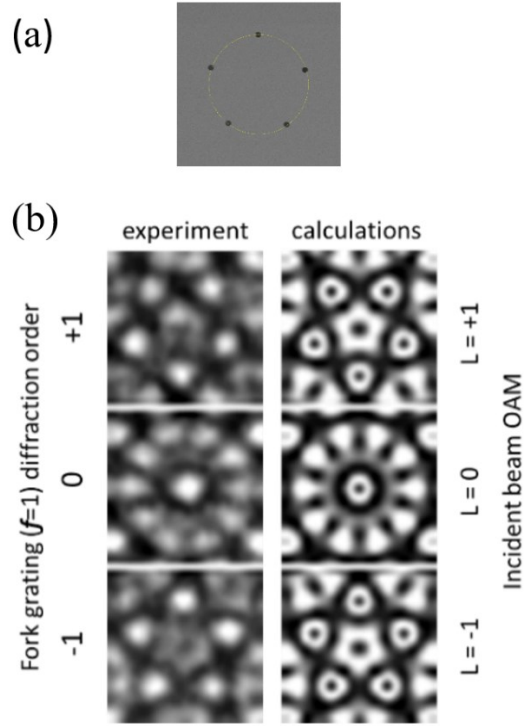

**Figure 7**

(a) SEM image of the opaque mask with five apertures at the vertices of a pentagon inscribed into a circle of diameter 20 μm. (b) Left column: experimental interference patterns obtained using the *n* = -1, 0, +1 orders of the FG (*f* = 1, *p* = 600 nm). Right column: calculated patterns for an incoming beam with **L** = -1, 0, +1.

## 5. OAM maths with SZP and FG devices

Here we examine the case of a FG being illuminated by a beam that already carries an OAM **L** ≠ 0, generated either by a SZP or by another FG, and show how one can establish an *integer arithmetic*s with OAM beams.

Fig. 8(a) illustrates the case of the incoming radiation produced by a SZP and carrying an **L** = *s* value of 0, +1 or -1 impinges on a FG with (*f* = 1), generating *n* = ±1 diffraction peaks. For an incoming **L** = 0 beam (middle image), the two rings corresponding to *n* = ±1 diffraction from the FG have the same size, associated to the expected **L** = ±1 values (see Fig. 7). Using a SZP with *s* = +1, imparting a value **L** = +1 to the incoming beam (bottom image), the *n* = +1 diffraction generates a wider ring [i.e., a larger L value according to Fig. 6(f)], while *n* = -1 features a closed spot instead of a ring, that we associate to an **L** = 0 beam. The opposite behavior is observed when using a SZP generating an incoming **L** = -1 beam (top image). These results show that the FG adds a value Δ**L** = *n* × *f* to the incoming beam, regardless of its initial **L** value.

The same approach can make use of two FGs in sequence, as shown in Fig. 8(b). Here, a first grating FG$_1$ with $f_1 = 1$ and $p_1 = 600$ nm is followed by a second grating FG$_2$ with $f_2 = 2$ and $p_2 = 250$ nm. When FG$_2$ is illuminated by the zero order of FG$_1$ ($n_1 = 0$), its $n_2 = \pm 1$ diffraction orders produce two equivalent rings, with expected **L** = ±2 values [Fig. 8(b), top]. Illuminating FG$_2$ with the $n_1 = -1$ beam (**L** = -1) results in a wider ring for $n_2 = -1$ and a narrower ring for $n_2 = +1$ [Fig. 8(b), bottom], which we interpret as the results of a sum of the OAM values imparted by each diffraction event, finally leading to a beam with **L** = $n_1 \times f_1 + n_2 \times f_2$.

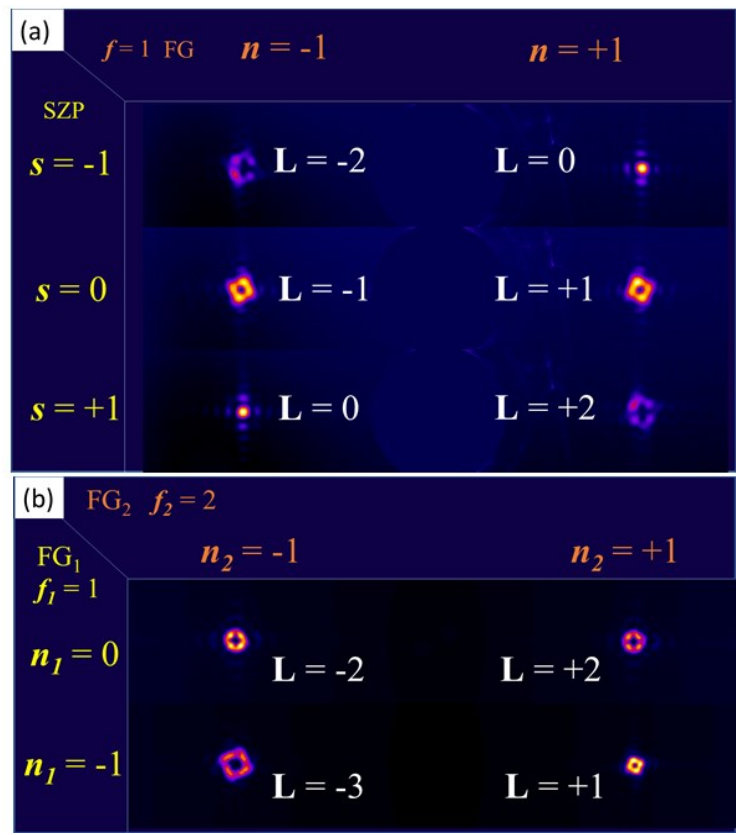

**Figure 8**

(a) Diffraction orders $n = \pm 1$ generated by the FG with $f = 1$ illuminated by a SZP with $s = -1$ (top), $s = 0$ (middle) and $s = +1$ (bottom). (b) Intensity pattern obtained by aligning two fork gratings FG$_1$ and FG$_2$ in cascade, the first one with $f_1 = 1$, the second with $f_2 = 2$. FG$_2$ is illuminated either by the zero order of FG$_1$ ($n_1 = 0$, top) or by its first negative order ($n_1 = -1$, bottom). In white, the expected **L** values for each spot at the detector.

Taken together, these two examples (i.e., SZP+FG and FG+FG) demonstrate that combining several devices (either SZPs or FGs) in cascade introduces a sort of OAM *integer arithmetic*, as it was already proposed in the visible range (Ruffato *et al.*, 2019). At the cost of introducing an additional diffractive element, *i.e.* at the expense of photon flux, OAM arithmetic adds flexibility to the control of the final **L** value of the resulting x-ray beam when starting from a limited number of FG or SZP devices.

It is also worth noting that the ensemble of results reported in Fig. 8 shows that a single FG device can be used for a simple, efficient and reliable test of the **L** value of the incoming x-ray beam, and we will make use of this capability in the next Section.

### 6. Harmonic emission from a helical undulator

Several years ago (Sasaki *et al.*, 2007; Sasaki & McNulty, 2008), it was predicted that the circularly polarized light produced by helical undulators (HUs) carries, in addition to the spin angular momentum of value **σ** (in $\hbar$ units) defined by the sign of the helicity (**σ** = ±1), an OAM of value **L** = **σ**(**h**-1), where **h** ≥1 is the harmonic order of the undulator emission (**h** = 1 for the fundamental emission). This was experimentally demonstrated at both synchrotron (Sasaki, 2007; Bahrdt *et al.*, 2013) and FEL (Hemsing, 2014; Rebernik Ribič *et al.*, 2017) sources. The characterization of the OAM value of the HU emission was mostly performed under out-of-focus conditions and with minimal intervention of optical elements. Here, we address this property at the experimental chamber positioned at the end of the beamline.

The SEXTANTS beamline is equipped with two HUs of different periods to cover the full energy range in fundamental emission; however, for radiation security reasons, the two devices cannot be operated simultaneously. Therefore, the classical test of detecting the OAM value based on the interference of same-wavelength emission (Bahrdt *et al.*, 2013) at, for example, **h**=1 and **h**=2 from distinct devices cannot be performed at SEXTANTS. Using the helical configuration of a single undulator (i.e. |**σ**|=1), we carried out three alternative tests based on the results shown in Sections 4 and 5, namely:

a) comparison, at the same photon energy and in circular polarization mode, of the intensity distributions for **h** = 1 and **h** = 2, expected to produce Gaussian and ring-shaped beams, respectively. These tests provided no clear evidence of ring-shaped intensity at or near the focal plane for **h** = 2.

b) comparison of the intensity distributions for **h** = 1 and **h** = 2 after scattering from the five-aperture mask of Fig. 7(a). In our measurements, we did not observe any change in the interference pattern comparable to that shown in Fig. 7(b).

c) comparison of the intensity patterns generated by a FG (*f* = 1, *n* = ±1) for **h** = 1 (expected **L**=0) and **h** = 2 (expected **L**=1) incoming circularly polarized beams. In this case as well, we did not observe any variation similar to those reported in Fig. 8(b), i.e. we had no clear indication of an incoming beam with **L** ≠ 0.

Summarizing, all checks performed so far at the level of the experimental chamber failed to demonstrate a well-defined **L** value for the **h** = 2 circularly polarized HU emission. As mentioned above, previous experiments attesting the OAM value in the HU emission were performed out of focus and with little or no intervention of optical elements. It is therefore likely that the well-defined **L** value of the diverging beam at the source is degraded during radiation transport and focusing along the beamline, where, at SEXTANTS, it encounters two slits, six mirrors and a grating before reaching its final destination at the IRMA-2 experimental chamber.

We conclude that, at present, we are not in position to propose the use of the radiation from the SEXTANTS HU sources for OAM-dependent experiments. Further tests will be carried out by the beamline staff, in collaboration with the Source and Optics groups at SOLEIL, to determine whether the beam transport and focusing conditions can be controlled and adapted to preserve the OAM of the x-rays down to the experimental stations.

## 7. Fourier transform holography with X-OAM beams

As a first pilot experiment with OAM x-ray beams at SEXTANTS, we performed imaging of magnetic materials via the resonant scattering of coherent x-rays, using the Fourier transform holography (FTH, Eisebitt *et al.*, 2004) method already implemented at the beamline (Sacchi *et al.*, 2012, Popescu *et al.*, 2019). This allows us to define better the conditions for future users' experiments, assessing the feasibility of element-selective magnetic imaging with x-rays carrying OAM in terms of setup features, beam characteristics, and capability of spanning a wide energy range.

We investigated two samples: the 50 nm thick $Co_{91}Tb_9$ alloy film already discussed in Section 3, and a $(Co_{1nm}/Gd_{0.5nm})\times 40$ multilayer deposited on a 100 nm thick $Si_3N_4$ membrane and FIB-etched to form a regular pattern of ~400 nm × 400 nm squares separated by ~100 nm wide lines. The CoGd sample had also its Au holography mask integrated on the opposite side of the $Si_3N_4$ membrane.

As shown in Section 5 and Fig. 6, FGs optimized for operation at the Gd and Tb 4d resonances also provide sufficient intensity at the Co-2p resonance. A holographic mask [Fig. 9(a)] was placed on top of the CoTb sample, and scattering patterns such as the one shown in Fig. 9(b) were collected as a function of photon energy (155 and 778 eV), light polarization (left/right circular) and **L** values (0, ±1), using the same FG with $f$ = 1 and $p$ = 600 nm. The magnetic image in Fig. 9(c) was obtained at the Co-2p resonance for **L** = +1, by taking the FT of the difference between scattering patterns acquired with opposite helicities. It is compared with a same-size MFM image [Fig. 9(d)] previously obtained from a different region of the same sample, showing a similar magnetic domain structure, but different details. Results obtained for **L** = ±1 at the Tb-4d (153eV) and Co-2p (778 eV) resonances are summarized and compared in Fig. 9(e-g).

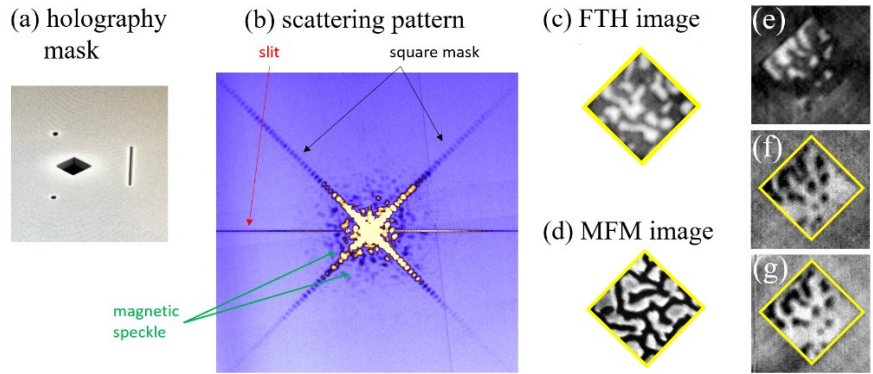

**Figure 9**

(a) Holography mask (1 µm Au) with 2.4 µm square aperture, two point-references and a slit, placed on top of the 50 nm thick $Co_{91}Tb_9$ sample. (b) Scattering pattern collected at 778 eV (Co-$L_3$ resonance) using a photon beam with right circular polarization and **L** = +1. Arrows associate different scattering features with sample and mask characteristics. (c) Image of the perpendicular magnetic domains obtained by taking the difference between circular left- and right-polarization measurements. (d) MFM image of the same sample acquired in a different region of the surface. (e-g) FTH images of magnetic domains in the $Co_{91}Tb_9$ sample obtained at the Tb 4d-4f resonance [(e), **L**=-1] and at the Co 2p-3d resonance [(f), **L**=-1, and (g), **L**=+1]. Diamonds indicate the size of the object aperture. The imaged sample area differs for (c) and (e-g).

We tested also the possibility of accessing the rare-earth 3d resonances in the 1170-1250 eV energy range. Fig. 10 shows an example of scattering through the 1.5 µm square holographic mask integrated into the $Si_3N_4$ substrate of the patterned Co/Gd multilayer sample. Both $N_{4,5}$ and $M_5$ Gd resonances were measured using the same FG optimized for 150 eV, changing only the undulator source and the CCD distance from the sample.

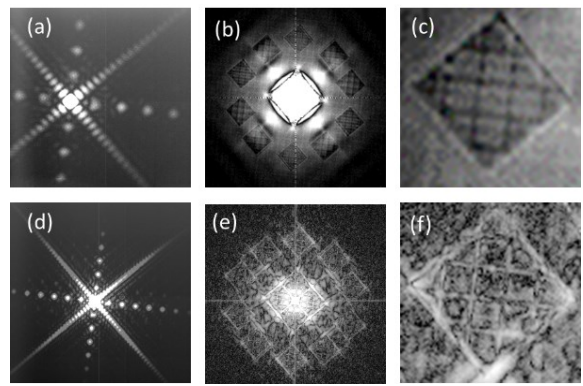

**Figure 10**

Scattering from the patterned CoGd multilayer measured at 150 eV [(a), Gd-$N_{4,5}$ resonance] and at 1185 eV [(d), Gd-$M_5$ resonance] using **L**=+1 radiation (FG with *p* = 600 nm and *f* = 1). (b, e): corresponding FTH images. (c, f): detail of an image of the object (1.5×1.5 µm²) produced by one of the references. The square pattern inside the object has a period of 500 nm.

The quality and intensity of the scattering signal, Figs. 10(a,d), are sufficiently good at both resonances; however, since this sample was not designed for operation at such high photon energies, the 1 μm thickness of the Au mask was not sufficient to suppress transmission at ~1200 eV. As a consequence, the FTH reconstructed images obtained at the Gd $M_5$ resonance are of significantly lower quality.

Overall, these results are very promising for performing **L**-dependent magnetic imaging experiments using holography and ptychography (Pancaldi *et al.*, 2024). The latter technique is more time-consuming because it requires raster scanning of the sample, but it is less demanding in terms of photon flux per image than holography, which relies on the intensity transmitted through the very small reference apertures.

Testing the use of OAM beams for x-ray FTH can be interesting for two reasons, related to the radial intensity dependence and azimuthal phase dependence of these beams. The former allows for a more favourable distribution of the intensity over the object and reference apertures, whose surface difference can easily exceed four orders of magnitude. Since FTH is based on the interference between beams scattered by the references and the object, placing the refences on the ring of maximum intensity and the object on the beam axis results in a better balance of their relative scattered amplitudes, beneficial for attaining a high-visibility interference pattern and, finally, images of better quality. The role of the azimuthal phase dependence in FTH is more difficult to assess: calculations predict an influence on the interference pattern, but this has not been verified experimentally yet, and additional work seems necessary to make use of this property of the OAM x-ray beams.

## 8. Conclusions

We have developed and implemented a new experimental setup within the IRMA-2 scattering chamber of the SEXTANTS beamline at the SOLEIL synchrotron. This setup enables the use of soft x-ray radiation carrying OAM, generated either by SZPs or by FGs, depending on the specific experimental requirements. Both transmission and reflection measurements can be performed with this setup.

Regarding the FGs, particular attention was paid to assessing their capability to cover a wide energy range using a single device in a single experimental configuration, without the need to intervene inside the vacuum vessel to replace or adjust optical elements. Scanning the photon energy of an OAM beam is especially attractive for spectroscopy experiments at synchrotron facilities.

We have performed test experiments based on resonant magnetic scattering (Figures 4, 11 and 12) and imaging by Fourier transform holography (Figures 9 and 10). We show that even a flux-demanding experiment such as magnetic imaging by FTH can be performed while spanning approximately one decade in photon energy. Our results demonstrate that, using a single FG and operating the standard beamline components (helical undulators, monochromator, focusing elements), it is possible to generate a soft x-ray beam with independently controlled spin and orbital angular momenta over a continuous and wide photon energy range. In particular, the 130-1200 eV range that we have explored covers most of the relevant resonances of multi-element transition-metal/rare-earth magnetic samples.

The combination of multiple SZP and/or FG devices in sequence has been proposed as a way of modifying the final **L** value. Together with the choice of the diffraction order for FGs, this introduces additional flexibility in the definition of the OAM of the final x-ray beam.

FGs have also been shown to be appropriate devices for a simple test of the OAM character of the incoming beam, providing an efficient diagnostic tool for synchrotron and free-electron laser x-ray sources.

The FGs used in these experiments are prototype devices. In particular, their limited area constrained their performance, which can be significantly improved through further optimization, as shown in Fig. 14. We are therefore confident in the technical feasibility of magnetic scattering and coherent imaging with soft x-ray OAM beams at the SEXTANTS beamline.

Finally, it is worth mentioning that the SOLEIL synchrotron is undertaking the process of a profound technical upgrade that will impact, among other aspects, the coherent flux of the source, which is a crucial parameter for optimizing the delivery of soft x-ray beams with controlled orbital angular momentum.

**Appendix A. Magnetic reflectivity with SZP-generated OAM beam**

Using the focused beam generated by the *s* = +1 SZP, we also performed high-angle reflectivity measurements, with the CCD detector mounted along the vertical axis. The main objective of this test was to assess the feasibility at SEXTANTS of using x-rays with controlled polarization and OAM for flux-demanding measurements, where reflectivities on the order of $10^{-6}$ are expected (~150 eV, p-polarization, angles close to the Brewster extinction condition). The sample was a $(Co_{1nm}/Gd_{0.5nm})\times 20$ multilayer deposited on a Si substrate, in which a pacman-shaped 18 μm dot has been isolated from the continuous film by FIB etching (Fig. 11, left). The 152 eV (Gd 4d resonance) x-ray beam was p-polarized (linear polarization oriented in the vertical scattering plane) and impinged at ~45° on the sample. The reflected intensity around 90° scattering angle (CCD images in Fig. 11) was measured for different values of the applied magnetic field and normalized to the average intensity over a full hysteresis cycle 20 mT wide. The large intensity disk (~6.3 mm in diameter) observed in each image corresponds to the SZP diffracted beam with **L** = +1, fully focused onto the Co/Gd dot and specularly reflected toward the CCD placed 450 mm from the sample. At the center of each image, a smaller spot corresponding to the **L**=0 beam transmitted through both the SZP and the square 0.5 mm OSA, is observed; its intensity is dominated by interaction with the continuous Co/Gd layer surrounding the 18 μm dot.

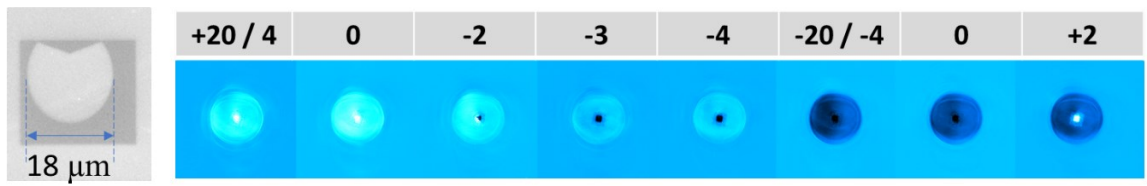

**Figure 11**

Normalized reflectivity from a 18 μm pacman-shaped Co/Gd dot (SEM image on the left), as a function of the applied magnetic field (in mT, above each image). Measurement conditions: SZP **L**=+1, OSA 0.5 mm, photon energy 152 eV (Gd-4d resonance), incidence angle 45°, linear p-polarization (vertical). Each image is 1024×1024 pixels, i.e. ~13.8mm wide.

The data show that, as a function of the applied field and in particular between 2 and 4 mT, the magnetization inside the dot behaves differently from that of the surrounding continuous film. Hysteresis curves obtained selecting the signal corresponding to the dot (large disk) and to the continuous film (small central spot) are compared in Fig. 12. Measurements of this type will enable users to envisage experiments on magnetic helicoidal dichroism similar to those already performed at free-electron laser sources, both in static (Fanciulli *et al.*, 2022) and pump-probe (Fanciulli *et al.*, 2025) modes.

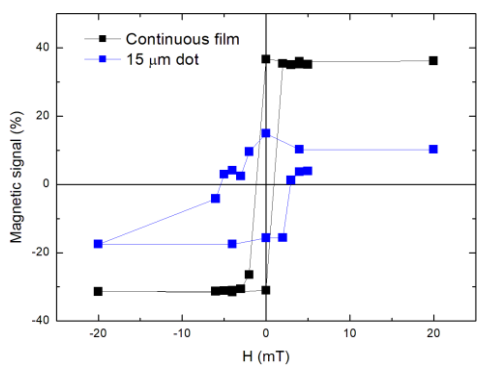

**Figure 12**

Hysteresis curves obtained selecting the magnetic signal corresponding to the 18 μm dot (large disk area, x-rays with **L**=1) and to the continuous film (small central spot, x-rays with **L**=0). The magnetic signal is defined as [I(H)-$I_{Ave}$]/$I_{Ave}$, where $I_{Ave}$ is the average signal over a complete cycle.

## Appendix B. First tests of larger FGs

A new set of FGs of larger size and optimized for working over the 600-1300 eV energy range has been fabricated at Institut Néel, Grenoble, and briefly tested at the SEXTANTS beamline. Eight FGs with $p$ = 200 nm or 400 nm and $f$ = 0, 1, 2, 3 are prepared on separate 200 nm thick $Si_3N_4$ membranes hosted on a single Si frame, according to the scheme shown in Fig. 13. Each FG is prepared by electron gun evaporation of 210 nm Au on a 60 μm × 60 μm square at the centre of each membrane, followed by electron beam lithography, Ti mask deposition and lift-off, and finally by ion-beam etching.

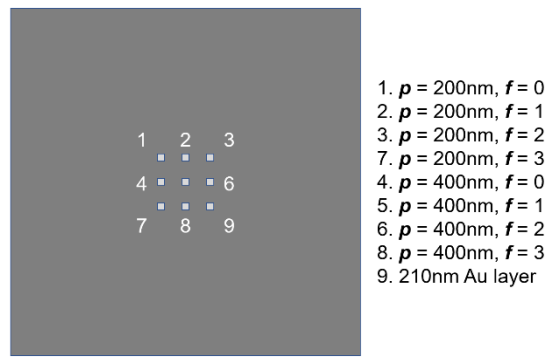

**Figure 13**
Sketch of a Si chip (7.5 mm wide) supporting nine $Si_3N_4$ square membranes (150 μm wide, 200 nm thick, 530 μm pitch), containing eight Au FGs and a continuous Au layer (210 nm thick). The area of each FG is 60×60 μm², their $p$ and $f$ values are detailed on the side.

The intensity measured at the 1st and 3rd diffraction orders using an x-ray beam ~40 × 30 μm² in size is shown in Fig. 14 (note the logarithmic scale), and compared with analogous measurements obtained using a 10 μm size FG (see Sec. 4). The strong diffraction from the edges observed for the latter is absent, or at least strongly reduced, for the larger 60 μm FG.

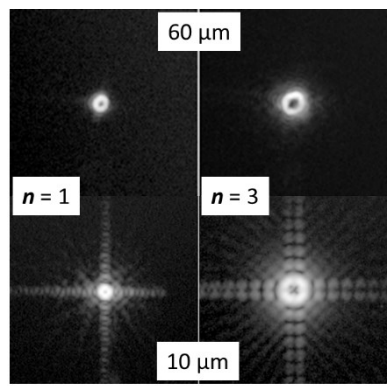

**Figure 14**
Diffraction from $f$ = 1 and $p$ = 200 nm square FGs with a lateral size of 60 μm (top) and 10 μm (bottom). The intensity of diffraction orders $n$ = 1 (left) and $n$ = 3 (right) is in logarithmic scale.

**Acknowledgements**   We acknowledge SOLEIL for provision of synchrotron radiation facilities and for assistance in using the SEXTANTS beamline under proposals n. 20211069, 20230032, 20231385 and 20250594. We would like to thank Thierry Crozes and Bruno Fernandez (Institut Néel, Grenoble) for their technical assistance with the cleanroom work.

**Funding information**   This research was supported by the Agence Nationale de la Recherche (France), project HELIMAG, Grant No. ANR-21-CE30-0037, and project TORNADO, Grant No. ANR-23-EXLU-0004/PEPR LUMA/France2030; the European Cooperation in Science and Technology (COST) Action NEXT, Grant No. CA22148.

**Conflicts of interest**   The authors declare no conflict of interest.

**Data availability**   Data supporting the results reported in this article are available upon reasonable request to the corresponding authors.

## References

Allen, L., Spreeuw, R. J. C., Woerdman, J. P. & Beijersbergen, M. W. (1992). Physical Review A. 45, 8185–8189.

Bahrdt, J., Scheer, M., Schmid, P., Müller, R., Kuske, P., & Holldack, K. (2013). First Observation of Photons Carrying Orbital Angular Momentum in Undulator Radiation. Physical Review Letters, 111(3). https://doi.org/10.1103/physrevlett.111.034801

Cojoc, D., Kaulich, B., Carpentiero, A., Cabrini, S., Businaro, L., & Di Fabrizio, E. (2006). X-ray vortices with high topological charge. Microelectronic Engineering, 83(4–9), 1360–1363. https://doi.org/10.1016/j.mee.2006.01.066

Daryabi, N., & Sabouri, S. G. (2023). Intersecting of circular apertures to measure integer and fractional topological charge of vortex beams. Optics Express, 31(17), 28459. https://doi.org/10.1364/oe.496425

Eisebitt, S., Lüning, J., Hellwig, O., Stöhr, J., Schlotter, W. F., Lörgen, M., & Eberhardt, W. (2004). Lensless imaging of magnetic nanostructures by X-ray spectro-holography. Nature, 432(7019), 885–888. https://doi.org/10.1038/nature03139

Fanciulli, M., Ruchon, T., Luttmann, M., Vimal, M., Sacchi, M., & Bresteau, D. (2021). Electromagnetic theory of helicoidal dichroism in reflection from magnetic structures. Physical Review A, 103(1). https://doi.org/10.1103/physreva.103.013501

Fanciulli, M., Capotondi, F., Bresteau, D., Pedersoli, E., Pancaldi, M., De Ninno, G., Rösner, B., Sousa, R., Prejbeanu, I.-L., Vila, L., Dieny, B., Ribič, P. R., Luttmann, M., Sacchi, M., De Angelis, D., Spezzani, C., Vimal, M., David, C., Ruchon, T., & Manfredda, M. (2022). Observation of Magnetic Helicoidal Dichroism with Extreme Ultraviolet Light Vortices. Physical Review Letters, 128(7). https://doi.org/10.1103/physrevlett.128.077401

Fanciulli, M., Carrara, P., Sacchi, M., Dieny, B., Pancaldi, M., David, C., De Angelis, D., De Ninno, G., Luttmann, M., Sousa, R., Pedersoli, E., Spezzani, C., Guer, M., Stanciu, A.-E., Manfredda, M., Rebernik Ribič, P., Capotondi, F., Prejbeanu, I. L., Buda-Prejbeanu, L. D., … Ravindran, A. (2025). Magnetic Vortex Dynamics Probed by Time-Resolved Magnetic Helicoidal Dichroism. Physical Review Letters, 134(15). https://doi.org/10.1103/physrevlett.134.156701

Fujita, H., & Sato, M. (2017). Encoding orbital angular momentum of light in magnets. Physical Review B, 96(6). https://doi.org/10.1103/physrevb.96.060407

Fürhapter, S., Ritsch-Marte, M., Bernet, S., & Jesacher, A. (2005). Spiral phase contrast imaging in microscopy. Optics Express, 13(3), 689. https://doi.org/10.1364/opex.13.000689

Gauthier, D., Ribič, P. R., Adhikary, G., Camper, A., Chappuis, C., Cucini, R., Dimauro, L. F., Dovillaire, G., Frassetto, F., Géneaux, R., Miotti, P., Poletto, L., Ressel, B., Spezzani, C., Stupar, M., Ruchon, T., & De Ninno, G. (2017). Tunable orbital angular momentum in high-harmonic generation. Nature Communications, 8(1). https://doi.org/10.1038/ncomms14971

Géneaux, R., Camper, A., Auguste, T., Gobert, O., Caillat, J., Taïeb, R., & Ruchon, T. (2016). Synthesis and characterization of attosecond light vortices in the extreme ultraviolet. Nature Communications, 7(1). https://doi.org/10.1038/ncomms12583

Hemsing, E., Hast, C., Dunning, M., Raubenheimer, T., & Xiang, D. (2014). First characterization of coherent optical vortices from harmonic undulator radiation. Physical Review Letters, 113(13). https://doi.org/10.1103/physrevlett.113.134803

Hickmann, J. M., Fonseca, E. J. S., & Jesus-Silva, A. J. (2011). Born's rule and the interference of photons with orbital angular momentum by a triangular slit. EPL (Europhysics Letters), 96(6), 64006. https://doi.org/10.1209/0295-5075/96/64006

Hu, H., Li, H., Xie, C., & Zhang, X. (2022). Self-standing quasi-random-dots fork gratings for single-order diffraction. Journal of Applied Physics, 132(22), 223105. https://doi.org/10.1063/5.0129269

Karimi, E., Piccirillo, B., Marrucci, L., Zito, G., & Santamato, E. (2007). Hypergeometric-Gaussian modes. Optics Letters, 32(21), 3053. https://doi.org/10.1364/ol.32.003053

Kohmura, Y., Mizumaki, M., Sawada, K., Ohwada, K., Ishikawa, T., & Watanuki, T. (2020). X-ray microscope for imaging topological charge and orbital angular momentum distribution formed by chirality. Optics Express, 28(16), 24115. https://doi.org/10.1364/oe.392135

Ladavac, K., & Grier, D. G. (2004). Microoptomechanical pumps assembled and driven by holographic optical vortex arrays. Optics Express, 12(6), 1144. https://doi.org/10.1364/opex.12.001144

Lee, J. C. T., Roy, S., Alexander, S. J., Mcmorran, B. J., & Kevan, S. D. (2019). Laguerre–Gauss and Hermite–Gauss soft X-ray states generated using diffractive optics. Nature Photonics, 13(3), 205–209. https://doi.org/10.1038/s41566-018-0328-8

Janicijevic, L., & Topuzoski, S. (2008). Fresnel and Fraunhofer diffraction of a Gaussian laser beam by fork-shaped gratings. Journal of the Optical Society of America A, 25(11), 2659. https://doi.org/10.1364/josaa.25.002659

Loetgering, L., Baluktsian, M., Keskinbora, K., Horstmeyer, R., Wilhein, T., Schütz, G., Eikema, K. S. E., & Witte, S. (2020). Generation and characterization of focused helical x-ray beams. Science Advances, 6(7). https://doi.org/10.1126/sciadv.aax8836

Luttmann, M., Fanciulli, M., Carrara, P., Sacchi, M., & Ruchon, T. (2025). Optical spin-orbit interaction induced by magnetic textures. https://doi.org/10.48550/arxiv.2506.15232

McCarter, M. R., Morley, S. A., Tremsin, A. S., Saleheen, A. I. U., De Long, L. E., Singh, A., Hastings, J. T., Roy, S., Woods, J. S., Scholl, A., & Tumbleson, R. (2023). Antiferromagnetic real-space configuration probed by dichroism in scattered x-ray beams with orbital angular momentum. Physical Review B, 107(6). https://doi.org/10.1103/physrevb.107.l060407

Moghaddasi Fereidani, R., Tang, Z., & Yong, H. (2025). Unveiling Molecular Symmetry Through Twisted X-Ray Diffraction. Chemphyschem : A European Journal of Chemical Physics and Physical Chemistry, 26(8). https://doi.org/10.1002/cphc.202401042

Nazirkar, N. P., Fohtung, E., Shi, X., Shi, J., & N'Gom, M. (2024). Coherent diffractive imaging with twisted X-rays: Principles, applications, and outlook. Applied Physics Reviews, 11(2). https://doi.org/10.1063/5.0179765

Pancaldi, M., Ravindran, A., Kourousias, G., Ruchon, T., Adams, D. E., David, C., Bykova, I., Sacchi, M., Simoncig, A., De Angelis, D., Bevis, C. S., De Ninno, G., Zangrando, M., Vavassori, P., Pedersoli, E., Barolak, J., Manfredda, M., Mancini, G. F., Guzzi, F., … Novinec, L. (2024). High-resolution ptychographic imaging at a seeded free-electron laser source using OAM beams. Optica, 11(3), 403. https://doi.org/10.1364/optica.509745

Popescu, H., Gaudemer, R., Medjoubi, K., Fortuna, F., Perron, J., Desjardins, K., Vacheresse, R., Jaouen, N., Pinty, V., Pilette, B., Luning, J., Sacchi, M., & Delaunay, R. (2019). COMET: a new end-station at SOLEIL for coherent magnetic scattering in transmission. Journal of Synchrotron Radiation, 26(Pt 1), 280–290. https://doi.org/10.1107/s1600577518016612

Ruchon, T., Fanciulli, M. & Sacchi, M. (2022). Magneto-Optics with light beams carrying orbital angular momentum, in *The 2022 Magneto-Optics Roadmap*, edited by A. Berger & P. Vavassori. Journal of Physics D: Applied Physics, 55(46), 463003. https://doi.org/10.1088/1361-6463/ac8da0

Rebernik Ribič, P., Mahne, N., Mincigrucci, R., David, C., Rösner, B., Spampinati, S., Principi, E., Manfredda, M., Masciovecchio, C., De Ninno, G., Allaria, E., Roussel, E., Simoncig, A., Foglia, L., Giannessi, L., Gauthier, D., Döring, F., & Mirian, N. (2017). Extreme-Ultraviolet Vortices from a Free-Electron Laser. Physical Review X, 7(3). https://doi.org/10.1103/physrevx.7.031036

Rouxel, J. R., David, C., Oppermann, M., Lacour, J., Mancini, G. F., Svetina, C., Diaz, A., Kinschel, D., Zinna, F., Chergui, M., Rösner, B., Cannelli, O., Bacellar, C., & Karpov, D. (2022). Hard X-ray helical dichroism of disordered molecular media. Nature Photonics, 16(8), 570–574. https://doi.org/10.1038/s41566-022-01022-x

Ruffato, G., Massari, M., & Romanato, F. (2019). Multiplication and division of the orbital angular momentum of light with diffractive transformation optics. Light, Science & Applications, 8(1). https://doi.org/10.1038/s41377-019-0222-2

Sacchi, M., Fortuna, F., Popescu, H., Tortarolo, M., Delaunay, R., Spezzani, C., & Jaouen, N. (2012). Magnetic imaging by Fourier transform holography using linearly polarized x-rays. Optics Express, 20(9), 9769. https://doi.org/10.1364/oe.20.009769

Sacchi, M., Maier, U., Gaudemer, R., Fortuna, F., Delaunay, R., Popescu, H., Avila, A., Spezzani, C., & Jaouen, N. (2013a). IRMA-2 at SOLEIL: a setup for magnetic and coherent scattering of polarized soft x-rays. Journal of Physics: Conference Series, 425(20), 202009. https://doi.org/10.1088/1742-6596/425/20/202009

Sacchi, M., Chiuzbaian, S. G., Hague, C. F., Cauchon, G., Dubuisson, J. M., Jaouen, N., Lagarde, B., Delmotte, A., Gaudemer, R., Popescu, H., Polack, F., & Tonnerre, J. M. (2013b). The SEXTANTS beamline at SOLEIL: a new facility for elastic, inelastic and coherent scattering of soft X-rays. Journal of Physics: Conference Series, 425(7), 072018. https://doi.org/10.1088/1742-6596/425/7/072018

Sanchez Brea, L. M. (2019) Python Diffraction and Interference. https://diffractio.readthedocs.io

Sasaki, S., Mcnulty, I., & Dejus, R. (2007). Undulator radiation carrying spin and orbital angular momentum. Nuclear Instruments and Methods in Physics Research Section A: Accelerators, Spectrometers, Detectors and Associated Equipment, 582(1), 43–46. https://doi.org/10.1016/j.nima.2007.08.058

Sasaki, S., & Mcnulty, I. (2008). Proposal for Generating Brilliant X-Ray Beams Carrying Orbital Angular Momentum. Physical Review Letters, 100(12). https://doi.org/10.1103/physrevlett.100.124801

Shen, Y., Min, C., Liu, Q., Xie, Z., Fu, X., Yuan, X., Gong, M., & Wang, X. (2019). Optical vortices 30 years on: OAM manipulation from topological charge to multiple singularities. Light: Science & Applications, 8(1). https://doi.org/10.1038/s41377-019-0194-2

Simpson, N. B., Allen, L., Dholakia, K., & Padgett, M. J. (1997). Mechanical equivalence of spin and orbital angular momentum of light: an optical spanner. Optics Letters, 22(1), 52. https://doi.org/10.1364/ol.22.000052

Tamburini, F., Anzolin, G., Barbieri, C., Bianchini, A., & Umbriaco, G. (2006). Overcoming the Rayleigh Criterion Limit with Optical Vortices. Physical Review Letters, 97(16). https://doi.org/10.1103/physrevlett.97.163903

Vila-Comamala, J., Sakdinawat, A., & Guizar-Sicairos, M. (2014). Characterization of x-ray phase vortices by ptychographic coherent diffractive imaging. Optics Letters, 39(18), 5281. https://doi.org/10.1364/ol.39.005281

Wang, B., Esashi, Y., Brooks, N. J., Jenkins, N. W., Murnane, M. M., Johnsen, P., Binnie, I., Tanksalvala, M., & Kapteyn, H. C. (2023). High-fidelity ptychographic imaging of highly periodic structures enabled by vortex high harmonic beams. Optica, 10(9), 1245. https://doi.org/10.1364/optica.498619

Woods, J. S., Kevan, S., Chopdekar, R. V., Kwok, W.-K., Chen, X. M., Hastings, J. T., Farmer, B., Hu, W., Wilkins, S., Roy, S., De Long, L. E., Mazzoli, C., Koch, R., Scholl, A., & Tremsin, A. S. (2021). Switchable X-Ray Orbital Angular Momentum from an Artificial Spin Ice. Physical Review Letters, 126(11). https://doi.org/10.1103/physrevlett.126.117201

Yan, J., & Geloni, G. (2023). Self-seeded free-electron lasers with orbital angular momentum. Advanced Photonics Nexus, 2(03). https://doi.org/10.1117/1.apn.2.3.036001